\documentclass[11pt]{article}

\usepackage{amsmath, amssymb, amsfonts, amsthm}
\usepackage{geometry}
\usepackage{graphicx}
\usepackage{hyperref}
\usepackage[numbers]{natbib}

\newtheorem{definition}{Definition}
\newtheorem{remark}{Remark}

\numberwithin{equation}{section}
\title{Lagrangian Rotating Contracting Structures}
\author{F.J.\ Beron-Vera\\ Department of Atmospheric Sciences\\ Rosenstiel School of Marine, Atmospheric \& Earth Science\\ University of Miami\\ Miami, Florida, USA\\ \href{mailto:fberon@miami.edu}{\texttt{fberon@miami.edu}}
}
\date{Started: March 30, 2026; this version: \today}

\begin{document}

\maketitle

\begin{abstract}
We identify materially defined regions in unsteady two-dimensional flows that combine finite-time contraction with elevated accumulated intrinsic rotation along trajectories, which we term \emph{Lagrangian rotating contracting structures} (LRCS). These regions are detected using existing objective diagnostics---the Lagrangian-averaged vorticity deviation (LAVD) together with direct tests of material contraction---without relying on the geometry of LAVD level sets.

In strongly deforming flows, LAVD maxima need not correspond to vortical regions or be enclosed by regular level sets, rendering geometry-based identification unreliable. Nevertheless, regions exhibiting inward spiraling motion and contraction can be extracted by combining LAVD with a contraction criterion.

Applications to atmospheric and oceanic flows show that such behavior arises both in twisted LAVD fields generated at submesoscales and in mesoscale flows where it is enhanced by inertial effects, with finite-time contraction providing the dynamical constraint that isolates materially organized regions with elevated intrinsic rotation.
\end{abstract}

\noindent\textbf{We identify materially organized regions in unsteady flows that combine elevated accumulated intrinsic rotation with finite-time contraction. LAVD alone may attain maxima unrelated to vortical structure and need not exhibit usable level-set geometry in strongly deforming flows. By selecting regions suggested by LAVD and enforcing contraction, we isolate structures that capture inward spiraling motion in a range of geophysical flows.}

\section{Introduction}

The identification of coherent structures in unsteady flows has been a central topic in dynamical systems and fluid mechanics. A prominent class of approaches is based on elliptic Lagrangian coherent structures (LCS) \cite{Haller-Beron-12, Haller-15, Haller-23}, which identify materially coherent vortices through objective (material-frame-indifferent) criteria.

Two principal realizations have emerged. Rotationally coherent vortices (RCV) \cite{Haller-etal-16} are material regions bounded by closed level sets of the Lagrangian-averaged vorticity deviation (LAVD), whose boundary elements experience equal accumulated intrinsic rotation over a finite time interval. Black-hole vortices (BHV), introduced in \cite{Haller-Beron-13, Haller-Beron-14} and in an earlier form in \cite{Beron-etal-13}, are material regions whose boundaries are closed curves along which all elements undergo the same finite-time stretching by a factor $p$. These curves arise as limit cycles of the associated $p$--line field, from a variational principle, and can be interpreted as null geodesics of a flow-induced Lorentzian metric. Instantaneous analogues have also been proposed \cite{Serra-Haller-16}.

Complementary notions of finite-time coherence have been developed in a probabilistic framework, notably through transfer operator methods and dynamic Laplacian constructions \cite{Dellnitz-etal-09, Froyland-etal-19, Froyland-26}, which identify coherent or almost-invariant sets based on minimal mixing or boundary growth.

Despite their differences, these approaches share a common focus: the identification of material regions that resist filamentation and undergo limited deformation over finite time intervals. In strongly deforming and multiscale flows, however, this requirement becomes restrictive. In \cite{Beron-etal-19-PNAS}, coherent Lagrangian swirls (CLS) were identified using LAVD-based diagnostics while allowing for boundaries with non-negligible filamentation and relaxed convexity constraints, showing that materially coherent regions associated with elevated intrinsic rotation may persist even when their boundaries depart significantly from near-convex geometry.

The present work departs from this viewpoint by abandoning geometric constraints and instead enforcing a material criterion. We consider regions that may undergo substantial deformation while remaining dynamically coherent through finite-time contraction.

In many geophysical flows, trajectories exhibit inward spiraling motion while the enclosing region contracts, even as its boundary stretches and filaments. Such behavior is not captured by instantaneous Eulerian diagnostics, including streamline patterns, which do not reflect material organization over time. This motivates focusing on properties that remain meaningful under strong deformation. Here we identify materially defined regions that combine finite-time contraction with elevated accumulated intrinsic rotation along trajectories over a finite time interval. These regions need not preserve their shape; rather, they are characterized by the coexistence of inward motion and intrinsic rotation. In this sense, they are more closely related to stable spirals in dissipative dynamical systems than to elliptic LCS. In an autonomous linear setting, such behavior is associated with complex eigenvalues $\lambda=\alpha\pm i\beta$ with $\alpha<0$, leading to rotation combined with contraction. The structures introduced here provide a finite-time, nonautonomous analogue of this behavior for extended material regions.

We do not introduce new diagnostics. Instead, we combine existing objective measures with material criteria. Accumulated intrinsic rotation is quantified by LAVD \cite{Haller-etal-16}, reviewed in Section~\ref{sec:setup}, while coherence is assessed through finite-time contraction of material regions (Section~\ref{sec:lrcs}). Elevated LAVD identifies trajectories with significant accumulated intrinsic rotational activity, but does not by itself indicate a vortical or materially organized region (Appendix~\ref{app:sourcesink}); high LAVD values may also arise in strongly deforming flows. The additional requirement of contraction provides the dynamical constraint needed to isolate physically meaningful structures.

We refer to material regions that exhibit both elevated intrinsic rotation, as quantified by LAVD, and finite-time contraction as \emph{Lagrangian rotating contracting structures} (LRCS). Their definition and detection are introduced in Section~\ref{sec:lrcs}. The behavior of the LAVD field in a strongly deforming atmospheric flow is examined in Section~\ref{sec:twisted}, and examples in geophysical flows, including a submesoscale oceanic flow and a mesoscale inertial flow, are presented in Section~\ref{sec:examples}.

\section{Setup}\label{sec:setup}

Let $D \subset \mathbb{R}^2$ be an open spatial domain, and let $\mathbf x = (x,y) \in D$, where $x$ and $y$ denote Cartesian coordinates. Consider a time-dependent velocity field $\mathbf u(\mathbf x,t) = (u(\mathbf x,t), v(\mathbf x,t))$. We assume that $\mathbf u$ is continuous in time and continuously differentiable in space, i.e., $\mathbf u \in C^1(D \times [t_0,t_1])$. No assumption of incompressibility is made, so that $\nabla\cdot\mathbf u$ may vary in space and time.

Trajectories $\mathbf x(t;\mathbf x_0,t_0)$ are defined as solutions of
\begin{align}
    \dot{\mathbf x} &= \mathbf u(\mathbf x,t), \quad \mathbf x(t_0) = \mathbf x_0,
\end{align}
and exist uniquely on $[t_0,t_1]$ for all initial conditions $\mathbf x_0 \in D$ whose trajectories remain in $D$. 

The associated flow map is
\begin{align}
    F_{t_0}^t : D &\to \mathbb{R}^2, \quad \mathbf x_0 \mapsto \mathbf x(t;\mathbf x_0,t_0),
\end{align}
which is continuously differentiable with respect to the initial condition $\mathbf x_0$. 

The scalar vorticity is defined by
\begin{align}
    \omega(\mathbf x,t) := \partial_x v(\mathbf x,t) - \partial_y u(\mathbf x,t).
\end{align}
Let $\bar\omega(t)$ denote the spatial mean of $\omega(\cdot,t)$ over $D$. The LAVD is defined as \cite{Haller-etal-16}
\begin{align}
    \mathrm{LAVD}_{t_0}^{t_1}(\mathbf x_0) := \int_{t_0}^{t_1} \bigl|\omega\bigl(F_{t_0}^t(\mathbf x_0),t\bigr) - \bar\omega(t)\bigr|\,dt.
\end{align}
The quantity $\mathrm{LAVD}_{t_0}^{t_1}$ is objective, i.e., invariant under time-dependent Euclidean changes of frame. In particular, subtraction of the spatial mean vorticity $\bar\omega(t)$ removes the contribution of rigid-body rotation, ensuring that $\mathrm{LAVD}_{t_0}^{t_1}$ measures intrinsic rotation relative to the bulk motion of the domain.

Throughout, we assume that trajectories initiating in the regions of interest remain in $D$ over $[t_0,t_1]$, so that all quantities above are well-defined.

We will be particularly interested in velocity fields in which material regions may undergo finite-time contraction while the trajectories they contain exhibit elevated accumulated intrinsic rotation over the interval considered. In such flows, elevated LAVD alone does not identify material coherence: it reflects accumulated intrinsic rotation along trajectories, which may also arise without coherent material organization. Accordingly, LAVD must be complemented by a finite-time contraction requirement to isolate materially coherent regions with elevated accumulated intrinsic rotation over the interval considered.

\section{Twisted LAVD in a compressible atmospheric flow}\label{sec:twisted}

Spiral cloud and precipitation patterns are a defining feature of tropical cyclones, reflecting the strongly rotating and convergent nature of the underlying atmospheric flow. Unlike more slowly evolving oceanic flows, the atmospheric velocity field is compressible and rapidly varying.

We analyze this behavior using 850~hPa winds during the passage of Hurricane Irma \cite{Cangialosi-etal-18}, obtained from the ECMWF fifth-generation reanalysis (ERA5) \cite{Hersbach-etal-20}. The analysis is initialized at $t_0 =$ 6 September 2017 03:00 and carried out over a finite-time interval $T = 36~\mathrm{h}$ on a domain centered on the storm.

The ERA5 velocity field is defined on $[125^\circ\mathrm{W},\,55^\circ\mathrm{W}] \times [10^\circ\mathrm{N},\,60^\circ\mathrm{N}]$, while the LAVD is computed over the subdomain $[85^\circ\mathrm{W},\,72^\circ\mathrm{W}] \times [14^\circ\mathrm{N},\,26^\circ\mathrm{N}]$, selected to cover the storm suggested by the inward, cyclonic instantaneous streamline pattern. This initial localization is not objective and is used only to focus the analysis; in principle, LAVD can be computed over a larger domain and all admissible contracting regions identified. While the presence of the storm is visually evident in the velocity field, the identification of the associated structure here is based on objective, material criteria and does not rely on this prior information.

\begin{figure}[t!]
    \centering
    \includegraphics[width=0.7\textwidth]{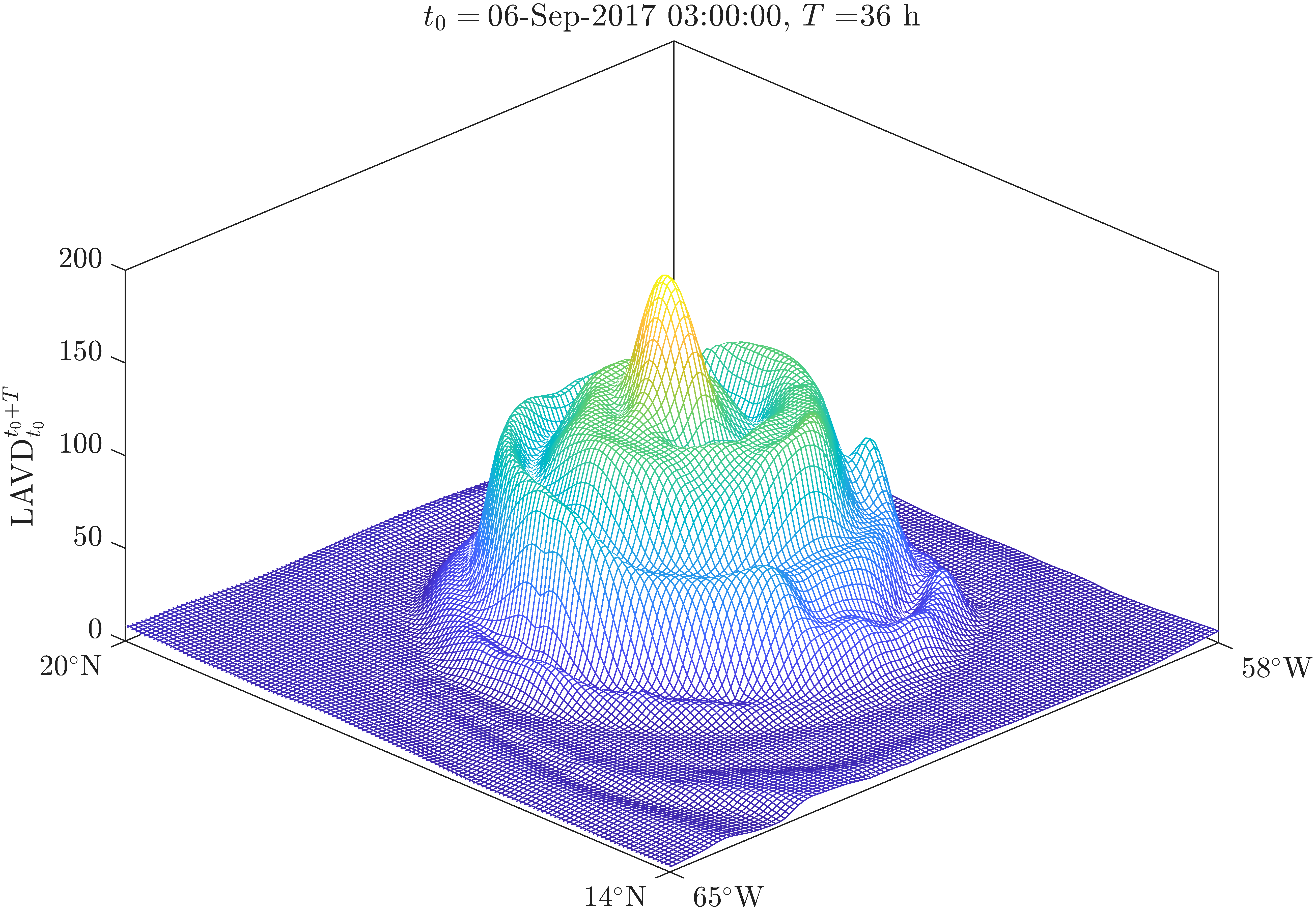}
    \caption{Surface representation of the LAVD field for the 850~hPa flow during Hurricane Irma over a $36$~h interval. A dominant central peak is surrounded by a twisted, asymmetric ridge structure. Mild smoothing has been applied; without it, multiple secondary maxima appear along the ridge.}
    \label{fig:irma_lavd_mesh}
\end{figure}

The resulting LAVD field departs markedly from the near-axisymmetric, smoothly nested structures often observed in flows with weak deformation or near-uniform rotational organization. As shown in Fig.~\ref{fig:irma_lavd_mesh}, a dominant central peak is embedded within a highly distorted ridge whose level sets are neither convex nor smoothly nested, but instead form a twisted, uneven elevation. Even after mild smoothing, residual undulations persist along the ridge; without smoothing, these appear as multiple secondary peaks, indicating that accumulated intrinsic rotation is distributed irregularly around the central maximum.

These features limit geometric interpretations based solely on level-set shape. Elevated LAVD identifies material subsets that share the same accumulated intrinsic rotation over the interval, but does not, by itself, distinguish material coherence. This motivates complementing LAVD with an explicit material requirement. In particular, combining elevated LAVD with a finite-time contraction criterion provides a basis for identifying material sets with elevated accumulated intrinsic rotation that remain coherent over the interval considered. We formalize this criterion in the following section.

\section{Lagrangian rotating contracting structures}\label{sec:lrcs}

\subsection{Definition}\label{sec:def}

\begin{definition}
A \textbf{Lagrangian rotating contracting structure} (LRCS) over a time interval $[t_0,t_1]$ is a material region $U(t)=F_{t_0}^t\bigl(U(t_0)\bigr)$, bounded by a closed material curve $\Gamma(t)=F_{t_0}^t\bigl(\Gamma(t_0)\bigr)$, such that:

(i) the region undergoes net contraction over $[t_0,t_1]$, i.e.,
\begin{align}
    |U(t_1)| < |U(t_0)|,
\end{align}

(ii) the initial region $U(t_0)$ contains a spatially localized region of elevated $\mathrm{LAVD}_{t_0}^{t_1}$.
\end{definition}

This definition combines a dynamical condition (finite-time contraction) with an objective measure of intrinsic rotation.

\begin{remark}
No geometric restriction is imposed on the boundary $\Gamma(t_0)$. In particular, admissible boundaries may exhibit significant geometric complexity, including loss of convexity and filamentary extensions under advection, while still enclosing a region that contracts over the time interval.
\end{remark}

\begin{remark}
LRCS are material regions, but their identification is boundary-based: in practice, closed material curves are extracted and tested, and the selected curve defines the LRCS as its interior.
\end{remark}

\begin{remark}
Elevated LAVD alone does not imply coherent material rotation (Appendix~\ref{app:sourcesink}). The contraction condition is therefore necessary to distinguish materially coherent rotating structures from regions of accumulated rotation in strongly deforming flows.
\end{remark}

\begin{remark}
The term ``rotating'' refers to intrinsic, trajectory-wise rotation accumulated over the time interval, not to rigid-body or bulk rotation of the region as a whole.
\end{remark}

\subsection{Detection procedure}\label{sec:detection}

Given a velocity field over $[t_0,t_1]$, LRCS are identified as follows:

\begin{enumerate}
\item Compute the scalar field $\mathrm{LAVD}_{t_0}^{t_1}(\mathbf x_0)$ over the initial domain.

\item Identify regions of elevated $\mathrm{LAVD}_{t_0}^{t_1}$, for instance via local maxima and their surrounding level sets or superlevel sets. Closed curves enclosing such regions are taken as candidates $\Gamma(t_0)$.

\item Retain only geometrically admissible candidates (e.g., closed curves of sufficient resolution and enclosed area), excluding trivial or poorly resolved contours.

\item For each candidate $\Gamma(t_0)$, let $U(t_0)$ denote its interior and define the normalized excess
\begin{align}
    \mathcal{E}(\Gamma) := \frac{1}{|U(t_0)|}
    \int_{U(t_0)}\bigl(\mathrm{LAVD}_{t_0}^{t_1}(\mathbf x_0)-c\bigr)_+\,d^2x,
\end{align}
where $c$ is the value of $\mathrm{LAVD}_{t_0}^{t_1}$ along $\Gamma(t_0)$.

\item Advect each candidate boundary to time $t_1$ and retain only those for which the enclosed region contracts, i.e.,
\begin{align}
    |U(t_1)| < |U(t_0)|.
\end{align}

\item Among contracting candidates, select those with large $\mathcal{E}(\Gamma)$, which quantify how strongly elevated LAVD is within the enclosed region. To avoid selecting contours that extend into strain-dominated surroundings and may undergo excessive filamentation, a mild inward buffering may be applied: rather than the outermost maximizing contour, a nearby interior level set is retained, yielding a more robust representation of the contracting, rotating core.
\end{enumerate}

In this procedure, LAVD serves to localize regions of elevated intrinsic rotation, while finite-time contraction provides the defining criterion for material coherence.

\section{Examples}\label{sec:examples}

We illustrate the identification of LRCS in three geophysical flows exhibiting distinct LAVD geometries. In all cases, regions combining elevated accumulated intrinsic rotation with finite-time contraction are identified, independently of the geometric properties of LAVD level sets. The examples include a strongly distorted atmospheric flow (Hurricane Irma), a high-resolution submesoscale-permitting oceanic flow (horizontal scales $\sim 0.1$--$10$~km) in which twisting and contraction arise directly from the resolved dynamics, and, as a final case, a mesoscale ocean flow (horizontal scales $\sim 10$--$200$~km) in which such behavior is enhanced by finite-size (inertial) effects.

\subsection{Hurricane Irma: distorted LAVD and contraction}\label{sec:irma}

\begin{figure}[t!]
    \centering
    \includegraphics[width=.9\textwidth]{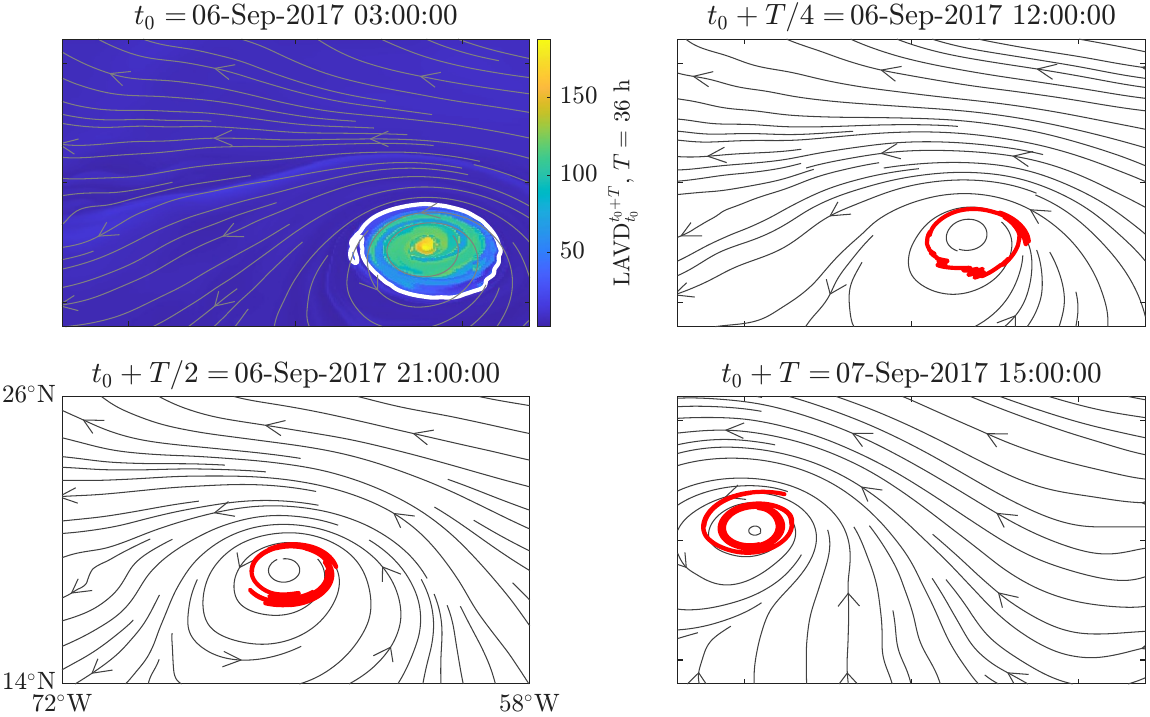}
    \caption{Evolution of the LRCS boundary for the 850~hPa flow during Hurricane Irma. The boundary is shown at $t_0$ (white) on the LAVD field and at later times (red) under advection, overlaid on instantaneous streamlines. The boundary undergoes substantial deformation while remaining organized around the storm center, and the enclosed region exhibits net contraction.}
    \label{fig:irma_lrcs}
\end{figure}

We revisit the Hurricane Irma flow introduced above. The LAVD field is strongly distorted and does not exhibit nested, regular level sets suitable for geometric selection of a material region. Figure~\ref{fig:irma_lrcs} shows the extracted boundary at $t_0$ (white) and its evolution (red).

The selected boundary, obtained by applying the procedure described above, encloses a region of elevated LAVD. As seen in Fig.~\ref{fig:irma_lrcs}, the boundary deforms substantially under advection over $[t_0,t_1=t_0+T]$, yet remains organized around the storm center, and the enclosed area decreases over the interval, indicating finite-time contraction. This first application shows that distorted LAVD geometry does not preclude the identification of rotating contracting regions: while LAVD provides candidate regions, the contraction criterion is necessary to isolate a materially consistent set.

\subsection{Submesoscale Gulf of Mexico flow: twisted LAVD and hidden spiraling}\label{sec:ncom}

Spiral patterns at the ocean surface have been documented in satellite imagery and in situ observations \cite{Munk-etal-00, DAsaro-etal-18, Beron-etal-15, Brach-etal-18}, but their relation to the underlying velocity field is not always evident, particularly when inferred from instantaneous diagnostics. We consider a surface velocity field from a high-resolution (1~km) NCOM (Navy Coastal Ocean Model) simulation of the Gulf of Mexico \cite{Jacobs-etal-14}, previously used in \cite{Beron-Lacasce-16, Beron-etal-19-PNAS}. The velocity field is defined on $[-97.95^\circ,-96^\circ] \times [22^\circ,26^\circ]$, with LAVD computed over the subdomain $[-97.3^\circ,-97^\circ] \times [24.1^\circ,24.6^\circ]$, selected to cover a region of positive vorticity suggestive of cyclonic motion. This initial localization is not objective and is used here only to focus the analysis; in principle, LAVD can be computed over a larger domain and all admissible contracting regions identified. Accordingly, the search for submesoscale LRCS in this dataset is not exhaustive, but illustrative.

\begin{figure}[t!]
    \centering
    \includegraphics[width=0.9\textwidth]{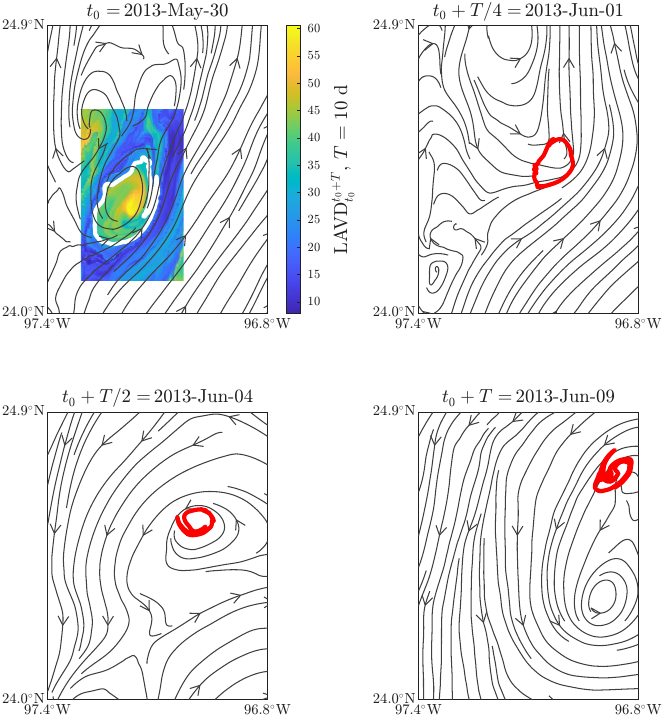}
    \caption{Submesoscale LRCS in a high-resolution NCOM surface flow. Top-left: LAVD field at $t_0$ with the extracted boundary (white). Remaining panels: advected boundary (red) overlaid on instantaneous streamlines at increasing times. The boundary contracts while undergoing inward spiraling. Instantaneous streamlines do not reveal this behavior.}
    \label{fig:ncom_twisted}
\end{figure}

The LAVD field exhibits a strongly twisted structure, with a localized maximum embedded within a distorted region of elevated values. Instantaneous streamlines at $t_0$ suggest a sink-like configuration, but this assessment is non-objective and does not reflect the material evolution of the flow. Figure~\ref{fig:ncom_twisted} shows the selected boundary at $t_0$ (white) and its evolution (red). The initial scale (mean radius) of the cyclonic LRCS is $\sim 10$--$20\,\mathrm{km}$ (with $1^\circ$ of latitude corresponding to approximately $111\,\mathrm{km}$ and longitude scales reduced by a factor $\cos\vartheta$ at latitude $\vartheta$).

As seen in Fig.~\ref{fig:ncom_twisted}, the boundary undergoes inward spiraling while exhibiting a net decrease in enclosed area, forming a tightly wound structure by $t_1=t_0+T$. Despite the absence of simple geometric organization in the LAVD field, the deformation remains coherent. Instantaneous streamlines do not reveal this behavior: although the flow appears rotational at fixed times, they do not capture the sustained inward spiraling and contraction revealed by the material evolution. This reflects the non-objective nature of streamline patterns and their inability to capture finite-time transport. This example demonstrates that twisted LAVD fields with contraction arise naturally in resolved submesoscale flows, and that instantaneous diagnostics may fail to reveal the associated rotating contracting regions.

\subsection{Mesoscale inertial Gulf Stream flow: regular LAVD with contraction}\label{sec:bom}

As a final example, we consider a velocity field constructed from altimetry-derived geostrophic flow \cite{LeTraon-etal-98}, augmented with a finite-size (inertial) correction that introduces weak compressibility. This correction follows from the Maxey--Riley equation \cite{Maxey-Riley-83} as adapted for particles floating at the ocean surface \cite{Beron-etal-19-PoF, Olascoaga-etal-20, Beron-21-ND}. In the long-time asymptotic limit for small particles, the dynamics reduce to a slow-manifold description, and the velocity field used here corresponds to a simplified form of this reduced model in the absence of wind forcing.

Under these conditions, the velocity can be written as
\begin{align}
    \mathbf u(\mathbf x,t)
    =
    \frac{g}{f_0}\,\nabla^\perp \eta(\mathbf x,t)
    +
    \tau g (1 - R)\,\nabla \eta(\mathbf x,t),
\end{align}
where $\eta(\mathbf x,t)$ denotes the sea-surface height, $g$ is the acceleration due to gravity, and $f_0$ is a reference Coriolis parameter; the operator $\nabla^\perp \eta = (-\partial_y \eta, \partial_x \eta)$ gives the geostrophic component. The second term represents the inertial correction, with $R$ a dimensionless parameter reflecting the particle-to-fluid density ratio (and air--water interaction) and $\tau$ the inertial response (Stokes) time, which depends on particle properties. The factor $\tau g(1-R)$ controls the strength of the ageostrophic contribution and hence the degree of compressibility.

In the present calculations, we set $\tau = 0.2~\mathrm{d}$ and $R = 0.9$, not to model a specific material, but to make the inertial contraction mechanism visually apparent \cite{Beron-21-ND}. The altimetry-derived velocity field is defined over a Gulf Stream domain $[70^\circ\mathrm{W},\,54^\circ\mathrm{W}] \times [33^\circ\mathrm{N},\,46^\circ\mathrm{N}]$, with analysis on the subdomain $[65^\circ\mathrm{W},\,61^\circ\mathrm{W}] \times [39^\circ\mathrm{N},\,41^\circ\mathrm{N}]$ over a time interval $T=30$ days. This subdomain was selected using the instantaneous streamline pattern, which suggests a cyclonic inward-spiraling feature. This localization is not objective and is used only to focus the example; the LRCS selection itself is based on the LAVD and contraction procedure described above.

\begin{figure}[t!]
    \centering
    \includegraphics[width=0.9\textwidth]{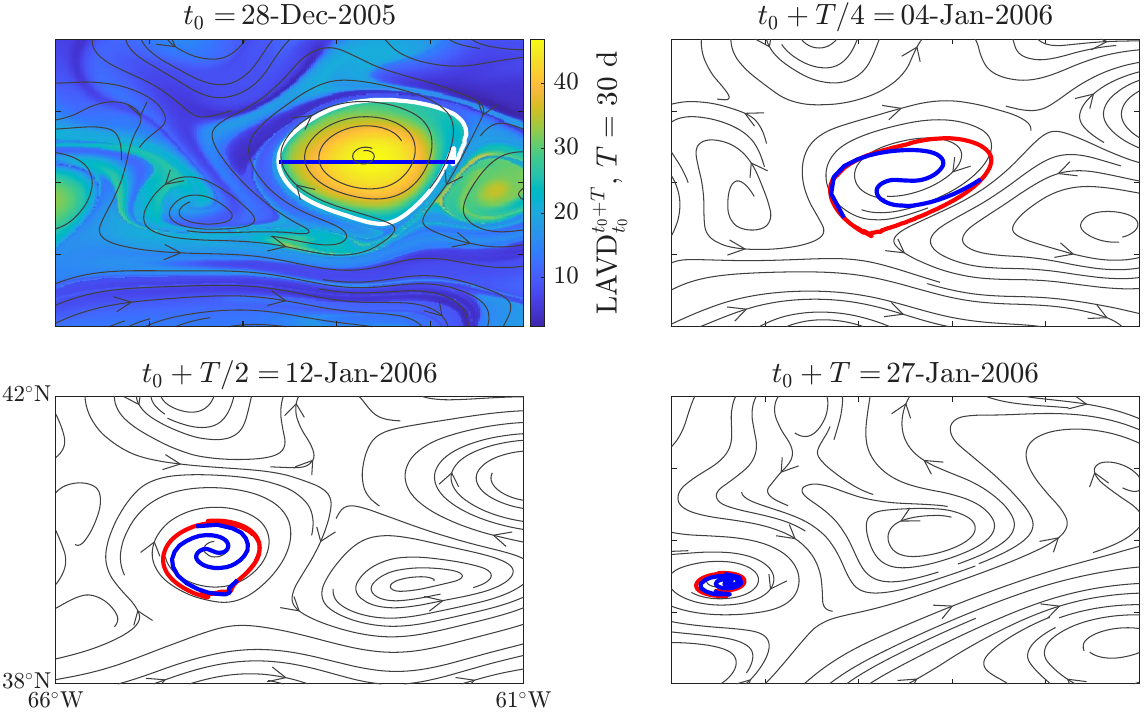}
    \caption{Evolution of the extracted boundary for the inertial Gulf Stream flow. The boundary is shown at $t_0$ (white) on the LAVD field and at later times (red) under advection, overlaid on instantaneous streamlines. The boundary undergoes net contraction with limited deformation.}
    \label{fig:ocean}
\end{figure}

The LAVD field exhibits a comparatively regular structure, with a dominant local maximum surrounded by approximately nested closed level sets, together with an additional local maximum that does not yield an admissible contracting boundary. Instantaneous streamlines also suggest a nearby anticyclonic source-like feature, but no corresponding LAVD maximum is present there. Figure~\ref{fig:ocean} shows the selected boundary at $t_0$ (white) and its evolution (red), from which it is seen that the boundary remains relatively regular while exhibiting a net decrease in enclosed area. Inward spiraling is not evident from the contracting boundary alone; it becomes apparent when following the evolution of a transversal material segment (blue), which reveals the winding motion within the region. This example shows that rotating contraction can arise even in flows with regular LAVD geometry, and that weak compressibility induced by inertial effects can enhance this behavior in mesoscale circulation.

\section{Conclusions}

We have introduced Lagrangian rotating contracting structures (LRCS) as materially defined regions in unsteady two-dimensional flows that combine finite-time contraction with elevated accumulated intrinsic rotation. The definition avoids geometric assumptions on diagnostic fields and instead relies on material behavior.

The approach does not introduce new diagnostics. It combines the Lagrangian-averaged vorticity deviation (LAVD), which measures accumulated intrinsic rotation along trajectories, with a direct test of finite-time contraction. Elevated LAVD identifies candidate regions but does not, by itself, indicate a vortical or materially organized structure. The contraction requirement provides a dynamical criterion to isolate such regions.

The examples illustrate this framework across different flow regimes. In the Hurricane Irma flow, the LAVD field is strongly distorted and lacks regular nested level sets, yet a rotating contracting region is identified through its material evolution. In a high-resolution NCOM simulation of the Gulf of Mexico, the LAVD field exhibits a twisted structure, and the selected region undergoes inward spiraling and contraction; this behavior is not evident from instantaneous streamlines. In the mesoscale inertial Gulf Stream flow, the LAVD field is more regular, but contraction remains necessary to identify the region, with inertial effects enhancing the rotating contracting behavior.

These examples indicate that neither the geometry of LAVD level sets nor instantaneous streamline patterns alone provide a reliable basis for identifying such structures. Rotating contraction may occur in flows with irregular, twisted, or regular LAVD fields and may not be apparent from instantaneous velocity diagnostics; it is established through the evolution of material regions.

The structures identified here can be viewed as a finite-time, nonautonomous analogue of stable spirals in dissipative dynamical systems. In this sense, the present work takes a step toward broadening the taxonomy of Lagrangian coherent structures to include materially defined regions characterized by finite-time contraction and elevated intrinsic rotation.

\section*{Acknowledgments}

This work was motivated by discussions within the International Space Science Institute (ISSI) team ``Opening new avenues in identifying coherent structures and transport barriers in magnetised solar plasma.''

\section*{Funding}

This work was carried out without any specific grant from funding agencies.

\section*{Author Declarations}

\subsection*{Conflict of Interest}

FJBV has no conflict of interest to disclose.

\subsection*{Author Contributions}

FJBV conducted all analyses and assumed full responsibility for the preparation and authorship of the manuscript.

\appendix

\section{LAVD maxima need not indicate material coherence}\label{app:sourcesink}

We illustrate a simple but instructive aspect of LAVD in a steady, two-dimensional compressible flow organized by alternating node and saddle structures. In this setting, streamlines are entirely source--sink driven and no vortical regions are present in any classical sense.

Consider the velocity field
\begin{align}
    u(x,y) = \sin\pi x,\quad
    v(x,y) = y(1-y^2)\cos\pi x.
\end{align}
This flow is steady and compressible, with alternating sources, sinks, and saddles organizing a repeating source--sink pattern. The discriminant $\Delta = (\mathrm{tr}\,\nabla \mathbf v)^2 - 4\det(\nabla \mathbf v)$ is nonnegative throughout the domain, confirming the absence of elliptic regions and hence of vortex-like flow geometry.

We also consider the instantaneous vorticity deviation (IVD), defined as the pointwise deviation of the vorticity from its spatial mean at a given time. As an objective instantaneous diagnostic, IVD highlights regions of locally elevated intrinsic rotation and serves as the Eulerian counterpart to LAVD.

\begin{figure}[t!]
    \centering
    \includegraphics[width=0.38\textwidth]{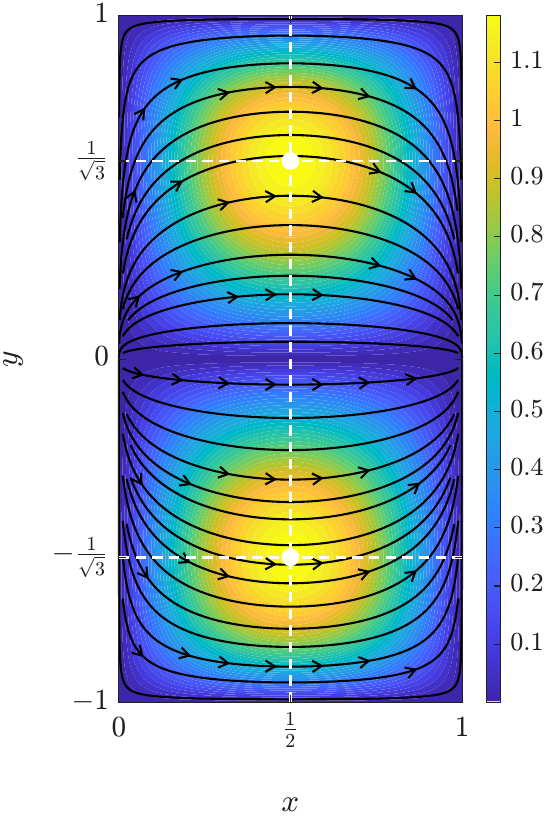}\,
    \includegraphics[width=0.58\textwidth]{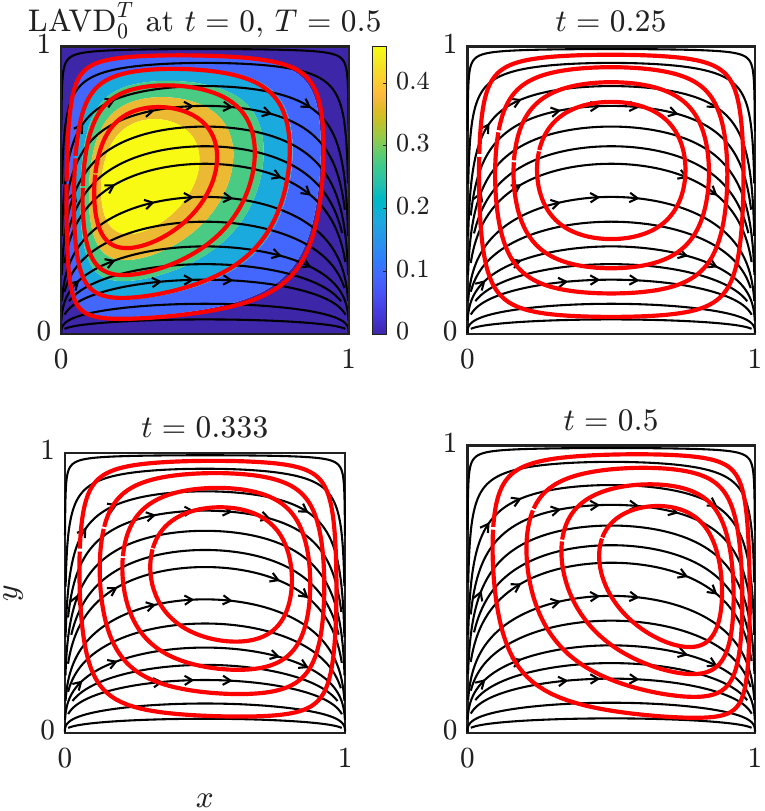}
    \caption{(left) IVD with streamlines. Closed IVD contours appear in regions between sources and sinks, despite the absence of closed streamlines or elliptic regions. (right) LAVD field with advected level sets. Closed material curves arise as level sets of accumulated intrinsic rotation, even though the flow has no vortex-like regions.}
    \label{fig:analytic_vd_streamlines}
\end{figure}

Figure~\ref{fig:analytic_vd_streamlines} shows the IVD and LAVD fields together with streamlines. Despite the absence of closed streamlines, both diagnostics exhibit organized structure, with maxima and closed level sets located in shear regions between sources and sinks. These structures do not correspond to vortex cores or coherent rotational motion, but instead reflect the accumulation of intrinsic rotation along trajectories.

Importantly, none of the LAVD level sets satisfies the defining contraction requirement of an LRCS. Any such level set chosen at $t_0$ encloses a region whose area is recovered at $t_1 = t_0+T$: the region expands away from the source, reaches a maximum extent approximately midway between source and sink, and subsequently contracts as trajectories approach the sink. Hence, no net contraction occurs over $[t_0,t_1]$ for any candidate boundary.

This example highlights a key point: elevated LAVD does not imply material coherence. Rather, LAVD organizes trajectories according to their accumulated intrinsic rotation, even in flows that lack vortex-like regions altogether.

Consequently, rotation diagnostics alone—even when objective—are insufficient to identify materially coherent sets. Additional dynamical criteria, most notably finite-time contraction, are required to isolate regions that behave as LRCS.

\section*{Data and Software Availability}

The gridded multimission altimeter products were produced by SSALTO/DUACS and distributed by AVISO (\href{https://\allowbreak www.aviso.altimetry.fr/}{{https://\allowbreak www.aviso.altimetry.fr/}}) with support from CNES. 

The wind velocity data originate from the European Centre for Medium-Range Weather Forecasts (ECMWF) Reanalysis v5 (ERA5), which can be accessed via \href{https://www.ecmwf.int/en/forecasts/dataset/ecmwf-reanalysis-v5}{https://www.ecmwf.int/en/forecasts/dataset/ecmwf-reanalysis-v5}.

The high-resolution NCOM (Navy Coastal Ocean Model) velocity fields used here are available upon reasonable request to the author. A comparable Gulf of Mexico NCOM simulation suitable for similar analyses is publicly available at \href{https://doi.org/10.7266/N7KH0K8G}{https://doi.org/10.7266/N7KH0K8G}.

The MATLAB scripts used to generate the figures in this paper are available from the corresponding author upon reasonable request. A generalized implementation in Julia is publicly available at \href{https://github.com/fberonvera/lrcs}{https://github.com/fberonvera/lrcs}. The Julia code is based on the author’s MATLAB implementations and was developed with the assistance of ChatGPT.

\bibliographystyle{alpha}
\newcommand{\etalchar}[1]{$^{#1}$}

\end{document}